\documentclass[prd,twocolumn,nofootinbib]{revtex4-1}
\usepackage{graphicx}
\usepackage{amssymb}
\usepackage{textcomp}
\usepackage{amsmath}
\usepackage{bm}
\usepackage{times}
\usepackage{epsfig}
\usepackage{color}
\usepackage{graphics}
\usepackage{hyperref}
\usepackage{setspace}

\hypersetup{
    pdfnewwindow=true,      
    colorlinks=true,       
    linkcolor=black,          
    citecolor=blue,        
    filecolor=blue,      
    urlcolor=blue           
}

\def\bea{\begin{eqnarray}}
\def\eea{\end{eqnarray}}

\begin{document}
\begin{flushright}
ULB-TH/17-11
\end{flushright}

\title{Displaced Photon Signal from a Light Scalar in Minimal Left-Right Symmetric Model}

\author{P. S. Bhupal Dev$^a$, Rabindra N. Mohapatra$^b$, Yongchao Zhang$^c$}
\address{$^a$Department of Physics and McDonnell Center for the Space Sciences, Washington University, St. Louis, MO 63130, USA\\
$^b$Maryland Center for Fundamental Physics, Department of Physics, University of Maryland, College Park, MD 20742, USA\\
$^c$Service de Physique Th\'{e}orique, Universit\'{e} Libre de Bruxelles, Boulevard du Triomphe, CP225, 1050 Brussels, Belgium}

\begin{abstract}
  We point out that in the minimal left-right realization of TeV scale seesaw for neutrino masses, the neutral scalar from the right-handed $SU(2)_R$ breaking sector could be much lighter than the right-handed scale. We discuss for the first time the constraints on this particle from low-energy flavor observables, find that the light scalar is necessarily long-lived. We show that it can be searched for at the LHC via displaced signals of a collimated photon jet, and can also be tested in current and future high-intensity experiments. In contrast to the unique diphoton signal (and associated jets) in the left-right case, a generic beyond Standard Model light scalar decays mostly to leptons or jets. Thus, the diphoton channel proposed here provides a new avenue to test the left-right framework and reveal the underlying neutrino mass generation mechanism.
\end{abstract}
\maketitle

{\section{Introduction}}
The discovery of neutrino masses has provided the first laboratory evidence for physics beyond the Standard Model (SM). The nature of the underlying new physics is however unclear and an ``all hands on deck" approach is called for to pinpoint this, since the result would have a profound impact on the ongoing new physics searches by narrowing the beyond SM landscape. We explore this question using the seesaw paradigm~\cite{seesaw} which is a simple and  well motivated way to understand neutrino masses, and considering its ultraviolet-complete realization within a TeV-scale left-right symmetric model (LRSM) framework~\cite{LR}, based on the gauge group ${\cal G}_{\rm LR} \equiv SU(2)_L\times SU(2)_R\times U(1)_{B-L}$.

The experimental signals of this  model have been extensively studied in the literature, and generally involve the heavy gauge bosons and heavy right-handed neutrinos (RHNs)~\cite{Keung:1983uu, Deppisch:2015qwa, Khachatryan:2014dka} or heavy Higgs bosons~\cite{Gunion:1986im, Dev:2016dja, Maiezza:2015lza, ATLAS:2014kca}. Here we propose a new complementary probe involving the LR symmetry breaking scalar sector, which is intimately related to the neutrino mass generation.

For the first time, we point out that the $SU(2)_R$ breaking scalar (denoted here by $H_3$) could be much lighter than the right-handed scale $v_R$.
Unlike the heavy-$H_3$ case, a light $H_3$ could be produced (off-shell) in e.g. $K$ and $B$ mesons, through its mixing with other scalars, and therefore its couplings are tightly constrained by the low-energy flavor changing neutral current (FCNC) data. In consequence, it decays mostly into two photons via the $SU(2)_R$ gauge interaction, which is suppressed by the right-handed scale $v_R$. This naturally pushes $H_3$ to be a long-lived particle, with the (Lorentz-boosted) decay length clearly dictated by $v_R$ and its mass. This is likely to be seen in high-intensity experiments, like SHiP and DUNE, and the high-energy collider LHC, in the latter it appears as a displaced vertex. The (displaced) photon signal could provide important information on the right-handed scale $v_R$ and the seesaw mechanism, in a way that is largely complementary to other probes of the LRSM. This is a specific feature of the LRSM that distinguishes it from other beyond SM light Higgs scenarios; for example in general models, a light scalar could mix with the SM Higgs and decay mostly into hadron jets and/or leptons. The (displaced) diphoton signal from light scalar decay could therefore be viewed, in some sense, as a ``smoking-gun'' signal of the LRSM.

\section{Light neutral scalar}
The minimal LRSM consists of the following Higgs fields:
\begin{eqnarray}
\Phi &=& \left(\begin{array}{cc}\phi^0_1 & \phi^+_2\\\phi^-_1 & \phi^0_2\end{array}\right), \quad
\Delta_R = \left(\begin{array}{cc}\Delta^+_R/\sqrt{2} & \Delta^{++}_R\\\Delta^0_R & -\Delta^+_R/\sqrt{2}\end{array}\right),
\label{eq:scalar}
\end{eqnarray}
which transform under ${\cal G}_{\rm LR}$ as ({\bf 2}, {\bf 2}, 0) and ({\bf 1}, {\bf 3}, 2), respectively. The group ${\cal G}_{\rm LR}$ is broken down to the EW gauge group by the triplet vacuum expectation value (VEV) $\langle \Delta^0_R \rangle = v_R$, whereas the EW symmetry is broken by the bidoublet VEV $\langle \Phi \rangle = {\rm diag}(\kappa,\kappa')$, with the EW VEV $v_{\rm EW} = \sqrt{\kappa^2 + \kappa^{\prime2}}$. For simplicity, we assume that the discrete parity symmetry has been broken at a scale much larger than the $SU(2)_R$-breaking scale~\cite{CMP}, but our conclusions remain unchanged in the TeV-scale fully parity-symmetric version of the LRSM.

The most general scalar potential involving $\Phi$ and $\Delta_R$ is given in Eq.~(\ref{eqn:potential}) in the appendix. 
One physical scalar from the bidoublet is identified as the SM Higgs $h$, while the other 4 degrees from the heavy doublet $(H_1, A_1, H_1^\pm)$  
have nearly degenerate mass, which is constrained to be $\gtrsim 10$ TeV from FCNC  constraints~\cite{Beall:1981ze}. Similarly, the mass of the doubly-charged scalars $H_2^{\pm\pm}$ from $\Delta_R$ is required to be above a few hundred GeV from same-sign dilepton pair searches at the LHC~\cite{ATLAS:2014kca}. However, no constraint is available in the literature for the remaining neutral scalar field $H_3$, consisting predominantly of the real component of $\Delta_R^0$. This is mainly due to the fact that it has no direct couplings to the SM sector and couples only to the heavy $SU(2)_R$ particles, in the limit of no mixing with other scalars. Therefore, its tree-level mass could in principle be much lower than the $v_R$ scale, as long as the quartic coupling $\rho_1 \ll 1$ [cf.~Eq.~(\ref{eqn:H30mass})]. This makes it the only possible light scalar in the model, and due to its suppressed couplings to the SM sector, it is also a natural LLP candidate at the LHC and in future colliders.

Since we envision that $H_3$ mass is much less than the $v_R$ scale, it is important to consider the loop corrections and see whether this small mass is radiatively stable. Recall that in the SM, if we neglect the one-loop fermion contributions to the Coleman-Weinberg effective potential~\cite{Coleman:1973jx}, there is a lower limit of order of 5 GeV on the Higgs boson mass~\cite{weinberg}. This bound goes away once the top-quark Yukawa coupling is included. Similarly, it was pointed out in Ref.~\cite{LH} that in a class of LRSM, there is a lower bound of about 900 GeV on the real part of the doublet scalar field coming from purely gauge contributions. 
Inclusion of the Yukawa interactions to the RHNs in the minimal LRSM we are considering allows us to avoid this bound and have a very light $H_3$.



Quantitatively, keeping only the $\Delta^0_R$ terms in the one-loop effective potential~\cite{RM}, we obtain the  correction term
\begin{eqnarray}
\label{eqn:loop}
\frac{3}{2\pi^2} \left[ \frac{1}{3} \alpha_3^2 + \frac83 \rho_2^2 - 8 f^4 + \frac{1}{2}g^4_R+(g^2_R+g^2_{BL})^2 \right] v^2_R \,,
\end{eqnarray}
where $g_R$ and $g_{BL}$ are respectively the $SU(2)_R$ and $U(1)_{B-L}$ gauge coupling strengths. We have assumed the three RHNs in the LRSM to be approximately degenerate with the same Yukawa coupling $f$.
From Eq.~\eqref{eqn:loop}, one would na\"ively expect the loop correction to be of order $v_R/4\pi$. However, the bosonic and fermionic contributions can cancel each other; with a mild tuning of $g_R$ and $f$ at the level of ${\rm GeV} / v_{\rm EW} \simeq 10^{-2}$, we can easily obtain a loop correction at or below the GeV scale. It is remarkable to note that the TeV scale seesaw prefers the natural value for Majorana Yukawa couplings to be of order one, implying  in turn TeV scale RHNs with observable same-sign dilepton plus dijet signatures at the LHC~\cite{Keung:1983uu}.



{\section{Couplings and decay}}
When the mass of $H_3$ is well below the EW scale, which is our focus in this letter, it decays to the light SM fermions through mixing with the SM Higgs $h$ and the heavy $CP$-even scalar $H_1$ from $\Phi$, with the mixing angles respectively given by
\begin{eqnarray}
\label{eqn:mixing}
\sin\theta_1 \ \simeq \ \frac{\alpha_1}{2\lambda_1}
\frac{v_R}{v_{\rm EW}} \,, \qquad
\sin\theta_2 \ \simeq \ \frac{4\alpha_2}{\alpha_3}
\frac{v_{\rm EW}}{v_R} \,.
\end{eqnarray}
Note the {\it inverted} dependence on the VEV ratio $(v_{\rm EW}/v_R)^{-1}$ for the $h-H_3$ mixing, because the SM Higgs boson mass is of order of $v_{\rm EW}$. 
The quartic couplings $\alpha_{1,2}$ connect $H_3$ to $h$ and $H_1$ respectively.  
There is an alignment limit of the parameter space for $\alpha_{1,2} \to 0$, when $H_3$ is secluded from mixing with other scalars in the LRSM, and $\lambda_1$ approaches to $\lambda_{\rm SM}=m_h^2/4v^2_{\rm EW}$. Thus for TeV-scale $v_R$, both the mixing angles $\sin\theta_{1,2}$ are {\it naturally} small.

At the one-loop level, the gauge and Yukawa couplings induce the decay of $H_3$ into digluons and diphotons, as in the SM Higgs case. However, when the FCNC constraints on the mixing angles $\sin\theta_{1,2}$ are considered (see below), the diphoton channel is dominated by the $W_R$ loop which is suppressed only by the RH scale $v_R$: $\Gamma_{\gamma\gamma} \propto v_R^{-2}$ but not sensitive to the gauge coupling $g_R$. The heavy charged scalar loops ($H_1^\pm$ and $H_2^{\pm\pm}$) are subleading, suppressed by a factor of $-5/21$ [cf.~Eq.~(\ref{eqn:gamma})]. The SM $W$ loop is heavily suppressed by the $W-W_R$ mixing. All the couplings and partial decay widths of $H_3$ are collected in the appendix.

\begin{figure}[!t]
  \centering
  \includegraphics[width=0.45\textwidth]{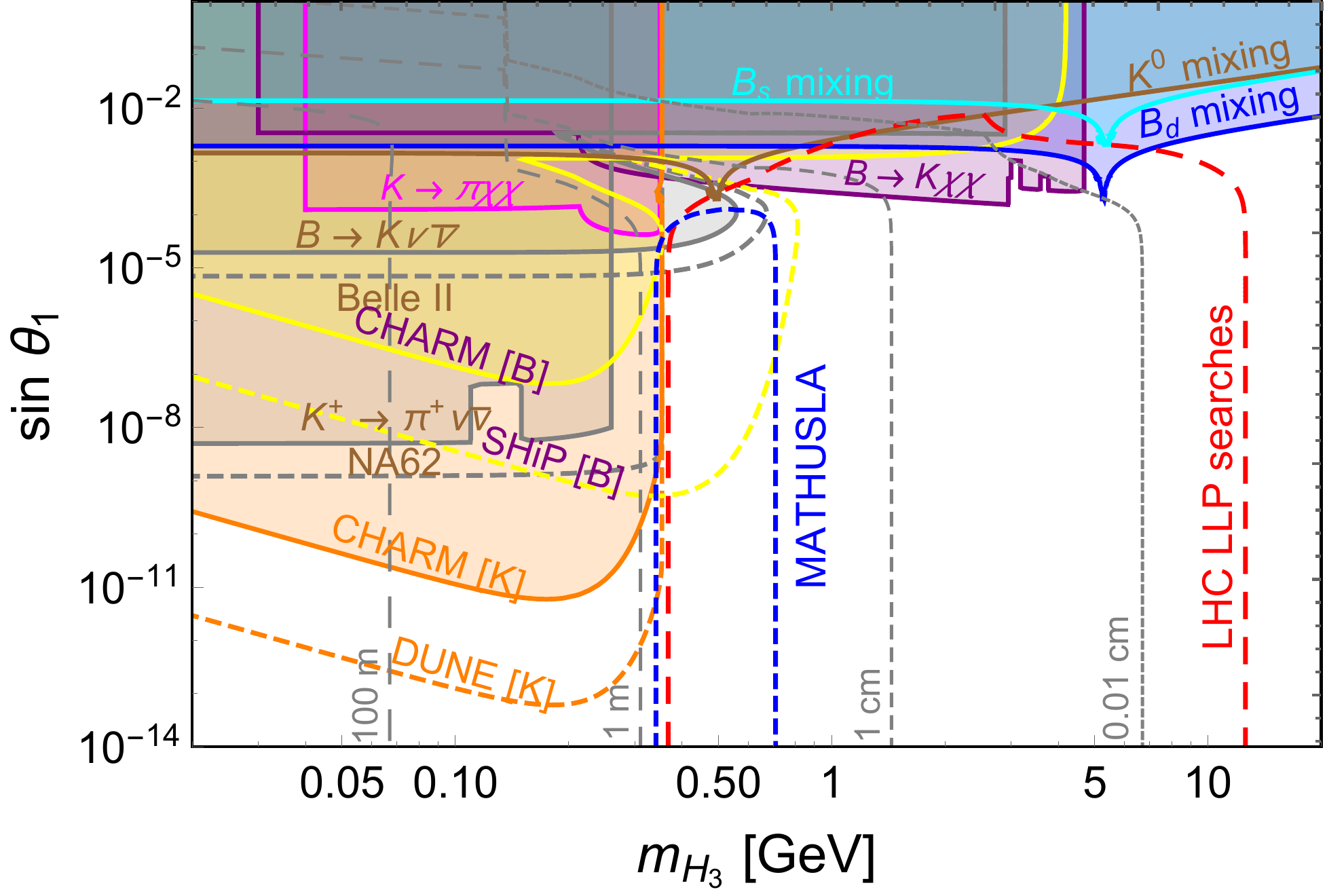}
  \caption{Contours of $H_3$ decay length at rest (dashed gray lines) as functions of its mass and mixing with the SM Higgs boson. Superimposed are limits (color-shaded) from meson mixing ($K^0$, $B_{d,s}$) and rare meson decays $K \to \pi \chi\chi$, $B \to K \chi\chi$ ($\chi = e,\, \mu,\, \gamma$), $K \to \pi \nu \bar\nu$, $B \to K \nu \bar\nu$ and $K \to \pi H_3 \to \pi \gamma\gamma$ and $B \to K H_3 \to K \gamma\gamma$ at beam-dump experiments. Also shown are the projected sensitivities from LLP searches at LHC and MATHUSLA.}
  \label{fig:limits}
\end{figure}

Contours of fixed decay length $L_0$ of $H_3$ {\it at rest} are shown in the $m_{H_3}-\sin\theta_1$ plane of Fig.~\ref{fig:limits} (dashed grey lines). For concreteness, we have made the following reasonable assumptions: (i) The RH scale $v_R = 5$ TeV, which is the smallest value required to satisfy the current LHC limits on $W_R$ mass. We also set the $H_3-H_1$ mixing $\sin\theta_2=0$. (ii) In the minimal LRSM, the RH quark mixing $V_R$ is very similar to the CKM matrix $V_L$, up to some additional phases~\cite{Senjanovic:2014pva}. For simplicity we adopt $V_R = V_L$ in the calculation. (iii) The couplings to charged leptons depend on the heavy and light neutrino sector via the Yukawa coupling matrix $Y_{\nu N}$. Here we assume the light neutrinos are of normal hierarchy with the lightest neutrino mass of 0.01 eV and the three RHNs degenerate at 1 TeV without any RH lepton mixing, which pushes the couplings $Y_{\nu N}\sim 10^{-7}$. Furthermore, the flavor-changing decay modes are included, such as $H_3 \to sb, \mu\tau$, and the running of strong coupling $\alpha_s$ is taken into consideration, which is important below the EW scale. 


From the lifetime curves in Fig.~\ref{fig:limits}, it is clear that when $m_{H_3}$ is below a few GeV, it tends to be long-lived, with decay lengths $L \gtrsim 0.01 b$ cm (where $b = E_{H_3} / m_{H_3}$ is the Lorentz boost factor, whose distribution typically peaks at around 100 for a GeV-scale $H_3$ produced at the LHC energy), as long as the mixing angles $\sin\theta_{1,2}$ are small $\lesssim 10^{-4}$, which is guaranteed by the flavor constraints, as discussed below. With the couplings to fermions constrained by the flavor data, only the diphoton channel is significant, implying that $H_3$ decays mostly into two {\it displaced} photons at the LHC.




We should mention here that, on the cosmological side,
when $H_3$ mass is below $\sim$50 MeV, it will start contributing to dark radiation as $\Delta N_{\rm eff} \simeq 4/7$, which is ruled out by the Planck data~\cite{Ade:2015xua} at the 2.5$\sigma$ C.L. Therefore, we will consider only $H_3$ with mass $\gtrsim 50$ MeV in the following.

{\section{Low-energy flavor constraints}}
Due to its mixing with the SM Higgs $h$ and the heavy scalar $H_1$, the light scalar $H_3$ induces flavor-changing couplings to the SM quarks, which are severely constrained by the low-energy flavor data, e.g. from $K-\bar{K}$, $B_d-\bar{B}_d$ and $B_s-\bar{B}_s$ neutral meson mixing, as well as rare $K$ and $B$ meson decays to lighter mesons and a photon pair. Although the couplings originate from the FCNC couplings of $H_1$, as the masses of $H_1$ and $H_3$ are independent observables, the flavor constraints on $H_3$ derived below  are different from those on the heavy scalar $H_1$~\cite{Beall:1981ze}.


Taking the $K^0 - \bar{K}^0$ mixing as an explicit example, we cast the flavor-changing four-fermion interactions mediated by $H_3$ into a linear combination of the effective dimension-6 operators of the form
\begin{align}
\mathcal{O} & \ = \
\mu_{RL}^2 \mathcal{O}_2 +
\mu_{LR}^2 \tilde{\mathcal{O}}_2
 + 2 \mu_{RL} \mu_{LR} \, \mathcal{O}_4 \, ,
\label{eqn:ope}
\end{align}
where $\mu_{RL,\,LR} = \sum_i m_i \lambda_i^{RL,\,LR}$ with $m_i = \{ m_u, m_c, m_t \}$ the running up-type quark masses, $\lambda_i^{LR} = V_{L,\, i2}^\ast V_{R,\, i1}$ and $\lambda_i^{RL} = V_{R,\, i2}^\ast V_{L,\, i1}$ the left- and right-handed quark mixing matrix elements, and
$\mathcal{O}_2 = (\bar{s} P_L d) (\bar{s} P_L d)$,
$\tilde{\mathcal{O}}_2 = (\bar{s} P_R d) (\bar{s} P_R d)$,
$\mathcal{O}_4 = (\bar{s} P_L d) (\bar{s} P_R d)$
with $P_{L,R} = \frac12 (1\mp\gamma_5)$~\cite{Babich:2006bh}.
The effective Lagrangian we need is thus given by
\begin{eqnarray}
\label{eqn:Leff}
\mathcal{L}_{H_3}^K & \ = \ & \frac{G_F}{\sqrt2}
\frac{\sin^2 \tilde{\theta}_2}{m_K^2 - m_{H_3}^2 + i m_{H_3} \Gamma_{H_3}}\mathcal{O} \, ,
\end{eqnarray}
where $G_F$ is the Fermi constant and $\sin\tilde\theta_2 = \sin\theta_2 + \xi \sin\theta_1$ is the ``effective'' mixing angle, which also involves the mixing with the SM Higgs, as $h$ mixes with $H_1$ with a small angle $\xi = \kappa'/\kappa \simeq m_b/m_t$~\cite{Dev:2016dja}. Although the flavor-changing couplings of $H_3$ arise from its mixing with $H_1$, the effective Lagrangian~\eqref{eqn:Leff} is not simply multiplied by a factor of $\sin\tilde\theta_2$; in particular, the operators of form $\mathcal{O}_2$ and $\tilde{\mathcal{O}}_2$ are absent in the $H_1$ case, which are canceled by the $CP$-odd scalar $A_1$ in the mass degenerate limit of $m_{H_1} = m_{A_1}$.
In Eq.~\eqref{eqn:ope}, the charm quark contribution $m_c \lambda$ dominates ($\lambda$ being the Cabibbo angle), with a subleading contribution $\sim m_t \lambda^5$ from the top quark.

Given the Lagrangian Eq.(5), it is straightforward to calculate the contribution of $H_3$ to the $K^0-\bar{K}^0$ mixing,
we need the hadronic matrix elements when the operators ${\cal O}_2$, $\tilde{\cal O}_2$ and ${\cal O}_4$ are sandwiched by the $K^0$ states,
\begin{eqnarray}
\langle K^0 | \mathcal{O}_i | \bar{K}^0 \rangle =
N_i m_K f_K^2 B_i (\mu) R_K^2 (\mu) \,,
\end{eqnarray}
with $i$ =2, 4, and $N_2 = 5/3$, $N_4 = -2$, $B_2 = 0.679$, $B_4 = 0.810$ from lattice calculation~\cite{Babich:2006bh} and the kaon decay constant $f_K = 113$ MeV. The mass ratio $R_K = m_K / (m_d + m_s)$ is evaluated at the energy scale $\mu=2$ GeV. As the strong interaction conserves parity, we have $\langle K^0 | \tilde{\mathcal{O}}_2 | \bar{K}^0 \rangle = \langle K^0 | \mathcal{O}_2 | \bar{K}^0 \rangle$. Then the $K^0$ mass difference
\begin{eqnarray}
\Delta m_K \simeq 2 \: {\rm Re} \, \eta_i (\mu)
\langle K^0 | \mathcal{L}^K_{H_3} | \bar{K}^0 \rangle \,,
\end{eqnarray}
with $\eta_2= 2.052$ and $\eta_4=3.2$ the the NLO QCD factors at $\mu=2$ GeV~\cite{Buras:2001ra}.

Requiring that the light $H_3$-mediated contribution be consistent with the current data on $\Delta m_K$, i.e. $<1.74 \times 10^{-12}$ MeV~\cite{PDG}, leads to an upper limit on the mixing angles $\sin {\theta}_{1,2}$, as presented in Fig.~\ref{fig:limits} (solid red line) for $\theta_1$ (the limit on $\theta_2$ is stronger by a factor of $\xi^{-1} \simeq m_t/m_b$).
As expected from the propagator structure in Eq.~\eqref{eqn:Leff}, the limits on the mixing angles $\sin\theta_{1,2}$ are significantly strengthened in the narrow resonance region where $m_{H_3} \simeq m_K$. For $m_{H_3}\ll m_K$, the $H_3$ propagator is dominated by the momentum term: $(q^2 - m_{H_3}^2 + i m_{H_3} \Gamma_{H_3})^{-1}\simeq q^{-2} \simeq m_K^{-2}$,
and the limit approaches to a constant value, whereas for $m_{H_3}\gg m_K$, the limit scales as $m_{H_3}$. 

The calculation of flavor constraints from $B_d$ and $B_s$ mixing are quite similar to those from $K^0$~\cite{next}. with the QCD correction coefficients $\eta_2 = 1.654$ and $\eta_4=2.254$~\cite{Buras:2001ra}, and the $B$-parameters
$B_2 (B_d) = 0.82$,
$B_4 (B_d) = 1.16$,
$B_2 (B_s) = 0.83$ and
$B_4 (B_s) = 1.17$~\cite{Becirevic:2001xt}. Unlike the $K^0$ case, the top-quark contribution dominates the effective coupling $\sum_i m_i \lambda_i^{LR,\, RL}$ and strengthens the corresponding limits on the couplings of $H_3$ to the bottom quark.
The mixing limits from $\Delta m_{B_d} < 9.3 \times 10^{-11}$ MeV and $\Delta m_{B_s} < 2.7 \times 10^{-9}$ MeV are shown in Fig.~\ref{fig:limits}, respectively, as the solid blue and cyan lines. The $B$ mesons are 10 times heavier than the $K$ meson, and the absolute values of error bars for $\Delta m_B$ are much larger than that for  $\Delta m_K$; this makes the $B$-mixing limits weaker than $K$-mixing limit for $m_{H_3} \ll m_B$. However, this could be partially compensated by the large effective coupling $\sum_i m_i \lambda_i^{LR}$ when $H_3$ is heavier. Thus for $m_{H_3} \gtrsim 1$ GeV, the limits on $\sin {\theta}_{1,2}$ from the $B_d$-mixing turn out to be more stringent. 

A light $H_3$ could also be produced in rare meson decays via the flavor-changing couplings, if kinematically allowed. The corresponding SM decay modes are either forbidden or highly suppressed by loop factors and the CKM matrix elements; thus these rare decay channels are also expected to set stringent limits on  $\sin\theta_{1,2}$. We consider the decays 
$B \to K H_3$ and $K \to \pi H_3$ each followed by $H_3\to \chi\chi$, with $\chi = e^+ e^-,\, \mu^+ \mu^-,\, \gamma\gamma$. The rare SM processes $K \to \pi \chi\chi$ and $B \to K \chi\chi$ has been searched for in NA48/2~\cite{Batley:2009aa, Batley:2011zz}, NA62~\cite{Ceccucci:2014oza}, KTeV~\cite{AlaviHarati:2003mr, AlaviHarati:2000hs, Abouzaid:2008xm, Alexopoulos:2004sx}, BaBar~\cite{Aubert:2003cm}, Belle~\cite{Wei:2009zv}, LHCb~\cite{Aaij:2012vr}. The limits on the mixing angle $\sin\theta_1$ are collectively depicted in Fig.~\ref{fig:limits}, where conservatively we demand $H_3$ decays inside the detector spatial resolution $L_{H_3} < 0.1$ mm, and the branching ratios ${\rm BR} (H_3 \to \chi\chi)$ and Lorentz boost factor $E_{H_3} / m_{H_3}$ from meson decay have been taken into consideration. More details can be found in Ref.~\cite{next}.

After being produced from meson decay, if $H_3$ decays outside the detector, the signal is $d_j \to d_i$  at the parton level plus missing energy. This could be constrained by the current limits of $K \to \pi \nu \bar\nu$ from E949~\cite{Anisimovsky:2004hr, Artamonov:2008qb, Artamonov:2009sz, Artamonov:2005cu} and $B \to K \nu \bar\nu$ from BaBar~\cite{Lees:2013kla}, and future prospects at  NA62~\cite{Anelli:2005ju} and Belle II~\cite{Abe:2010gxa}, which are all presented in Fig.~\ref{fig:limits}. As light $H_3$ tends to be long-lived, the ``'invisible'' searches with neutrinos in the final state are more constraining than ``visible'' decay modes above. With a huge number of protons-on-target and rather long decay length, the beam-dump experiments could further improve the limits. The current limits from CHARM~\cite{Bergsma:1985qz} and future prospects at SHiP~\cite{Alekhin:2015byh} and DUNE~\cite{Adams:2013qkq} are also shown in Fig.~\ref{fig:limits}, which could exclude the mixing angle up to the level of $10^{-13}$.

The full details of the limits of rare $K$ and $B$ decays on the couplings of $H_3$ are presented in Ref.~\cite{next}. Here we list only the most important information which leads to the limits and prospects in Fig.2. As $H_3$ can have tree-level flavor-changing couplings to the SM quarks, the decay $d_j \to d_i H_3$ in the down-type quark sector might exceed the observed total widths of $K$ and $B$ mesons, as long as the mixing angles $\sin\theta_{1,2}$ are sufficiently large. Thus in all the calculations below, we incorporate also the constraints of $\Gamma (K \to \pi H_3) > \Delta \Gamma_{\rm total}(K)$ and $\Gamma (B \to K H_3) > \Delta \Gamma_{\rm total}(B)$, where, taking into consideration of the theoretical and experimental uncertainties, we use 20\% of the total widths to set the limits. All the relevant rare decays
\begin{eqnarray}
d_j \to d_i H_3 \quad {\rm with} \quad
&& H_3 \to e^+ e^-,\, \mu^+ \mu^-,\, \nonumber \\
&& (\text{or}\; H_3 \to \text{any})
\end{eqnarray}
are collected in Table~\ref{tab:limits}, where the readers can find also the expected {\it average} energies of $H_3$ from meson decay and the current and future limits. For the ``visible'' decays with leptons or photons in the final state, if the decay length of $H_3$ is significantly larger than the detector spatial resolutions, the displaced events could easily be identified in the high intensity experiments, thus we conservatively set the decay length to be $L_{H_3} < 0.1$ mm, where the Lorentz boost factor $E_{H_3} / m_{H_3}$ has been taken into consideration. Regarding the $B$ decays, when the $H_3$ mass is close to that of $J/\psi$ or $\psi(2S)$, we use the SM branching ratios
\begin{eqnarray}
{\rm BR} (B \to K J/\psi) & \ = \ &
{\rm BR} (B \to K \ell^+ \ell^-) \ = \ 5 \times 10^{-5} \,, \nonumber \\
{\rm BR} (B \to K \psi (2S)) & \ = \ &
{\rm BR} (B \to K \ell^+ \ell^-) \ = \ 5 \times 10^{-6} \,. \nonumber \\
\end{eqnarray}
to set limits on $H_3$. For the ``invisible'' decays with neutrinos in the final state, $H_3$ is required to be long-lived enough to decay outside the detectors. In the beam-dump experiments CHARM, SHiP and DUNE, the most stringent limits are from the diphoton modes $H_3 \to \gamma\gamma$, benefiting from the large branching ratio. Without any signal observed, CHARM sets an upper limit of $N_{\rm event} < 2.3$ at the 90\% C.L., while at the future experiments SHiP and DUNE, we assume the signal numbers to be less than 3. More calculation details can be found in Ref.~\cite{next}.

\begin{table*}[!t]
  \centering
  \caption[]{Summary of meson decay constraints used to derive current/future limits in Fig.1. The last column gives the upper limit on the BR of the process used in our calculation. The corresponding numbers (in parenthesis) for the beam-dump experiments (last six rows) give the limit on the number of events. More details can be found in Ref.~\cite{next}.}
  \label{tab:limits}
  \begin{tabular}[t]{rlllll}
  \hline\hline
  Experiment & Meson decay & $H_3$ decay & $E_{H_3}$ & Decay length & Limit on BR ($N_{\rm event}$) \\ \hline
  NA48/2
  & $K^+ \to \pi^+ H_3$ & $H_3 \to e^+ e^-$ & $\sim 30$ GeV & $< 0.1$ mm &
  $2.63 \times 10^{-7}$ \\
  NA48/2
  & $K^+ \to \pi^+ H_3$ & $H_3 \to \mu^+ \mu^-$ & $\sim 30$ GeV & $< 0.1$ mm & $8.88\times10^{-8}$ \\
  NA62
  & $K^+ \to \pi^+ H_3$ & $H_3 \to \gamma\gamma$ & $\sim 37$ GeV & $< 0.1$ mm & $4.70\times10^{-7}$ \\ \hline
  E949
  & $K^+ \to \pi^+ H_3$ & any (inv.) & $\sim 355$ MeV & $> 4$ m & $4\times10^{-10}$ \\
  NA62
  & $K^+ \to \pi^+ H_3$ & any (inv.) & $\sim 37.5$ GeV & $> 2$ m & $2.4\times10^{-11}$ \\ \hline
  KTeV
  & $K_L \to \pi^0 H_3$ & $H_3 \to e^+ e^-$ & $\sim 30$ GeV & $< 0.1$ mm &
  $2.8 \times 10^{-10}$ \\
  KTeV
  & $K_L \to \pi^0 H_3$ & $H_3 \to \mu^+ \mu^-$ & $\sim 30$ GeV & $< 0.1$ mm & $4\times10^{-10}$ \\
  KTeV
  & $K_L \to \pi^0 H_3$ & $H_3 \to \gamma\gamma$ & $\sim 40$ GeV & $< 0.1$ mm & $3.71\times10^{-7}$ \\ \hline
  BaBar
  & $B \to K H_3$ & $H_3 \to \ell^+ \ell^-$ & $\sim m_B/2$ & $< 0.1$ mm & $7.91\times10^{-7}$ \\
  Belle
  & $B \to K H_3$ & $H_3 \to \ell^+ \ell^-$ & $\sim m_B/2$ & $< 0.1$ mm & $4.87\times10^{-7}$ \\
  LHCb
  & $B^+ \to K^+ H_3$ & $H_3 \to \mu^+\mu^-$ & $\sim 150$ GeV & $< 0.1$ mm & $4.61\times10^{-7}$ \\ \hline
  BaBar
  & $B \to K H_3$ & any (inv.) & $\sim m_B/2$ & $> 3.5$ m & $3.2\times10^{-5}$ \\
  Belle II
  & $B \to K H_3$ & any (inv.) & $\sim m_B/2$ & $> 3$ m & $4.1\times10^{-6}$ \\
  \hline
  CHARM
  & $K \to \pi H_3$ & $H_3 \to \gamma\gamma$ & $\sim 10$ GeV & $[480,\,515]$ m & $(< 2.3)$ \\
  CHARM
  & $B \to X_s H_3$ & $H_3 \to \gamma\gamma$ & $\sim 10$ GeV & $[480,\,515]$ m & $(< 2.3)$ \\
  SHiP
  & $B \to X_s H_3$ & $H_3 \to \gamma\gamma$ & $\sim 25$ GeV & $[70,\,125]$ m & $(< 3)$ \\
  DUNE
  & $K \to \pi H_3$ & $H_3 \to \gamma\gamma$ & $\sim 12$ GeV & $[500,\,507]$ m & $(< 3)$ \\
  \hline\hline
  \end{tabular}
\end{table*}

Note that the mixing angle $\sin\theta_1$ could also be constrained by the precise Higgs measurements, invisible SM Higgs decay, rare decays $Z \to \gamma H_3$ and $t \to u H_3,\, c H_3$. However, these limits are much weaker than those from meson oscillation and decay, at most of order $0.1$, and are not shown here.

{\section{Displaced diphoton signal at the LHC}}
For a light $H_3$ with mass $\lesssim 10$ GeV, the $h-H_3$ mixing is so severely constrained that its Higgs portal production is highly suppressed and it could only be produced via the gauge coupling through heavy vector boson fusion (VBF): $pp \to W_R^{\ast} W_R^\ast jj \to H_3 jj$, with a subleading contribution from $Z_R$ fusion~\cite{Dev:2016dja}. The associated production of $W_R H_3$ is further suppressed by the heavy gauge boson mass in the final state. When $m_{H_3} \lesssim 10$ GeV, the VBF production rate is almost constant for a given $v_R$, and is sensitive only to the gauge coupling $g_R$. For a smaller $g_R < g_L$, the $W_R$ boson is lighter and the production of $H_3$ can be significantly enhanced. 

\begin{figure}[!t]
  \centering
  \includegraphics[width=0.4\textwidth]{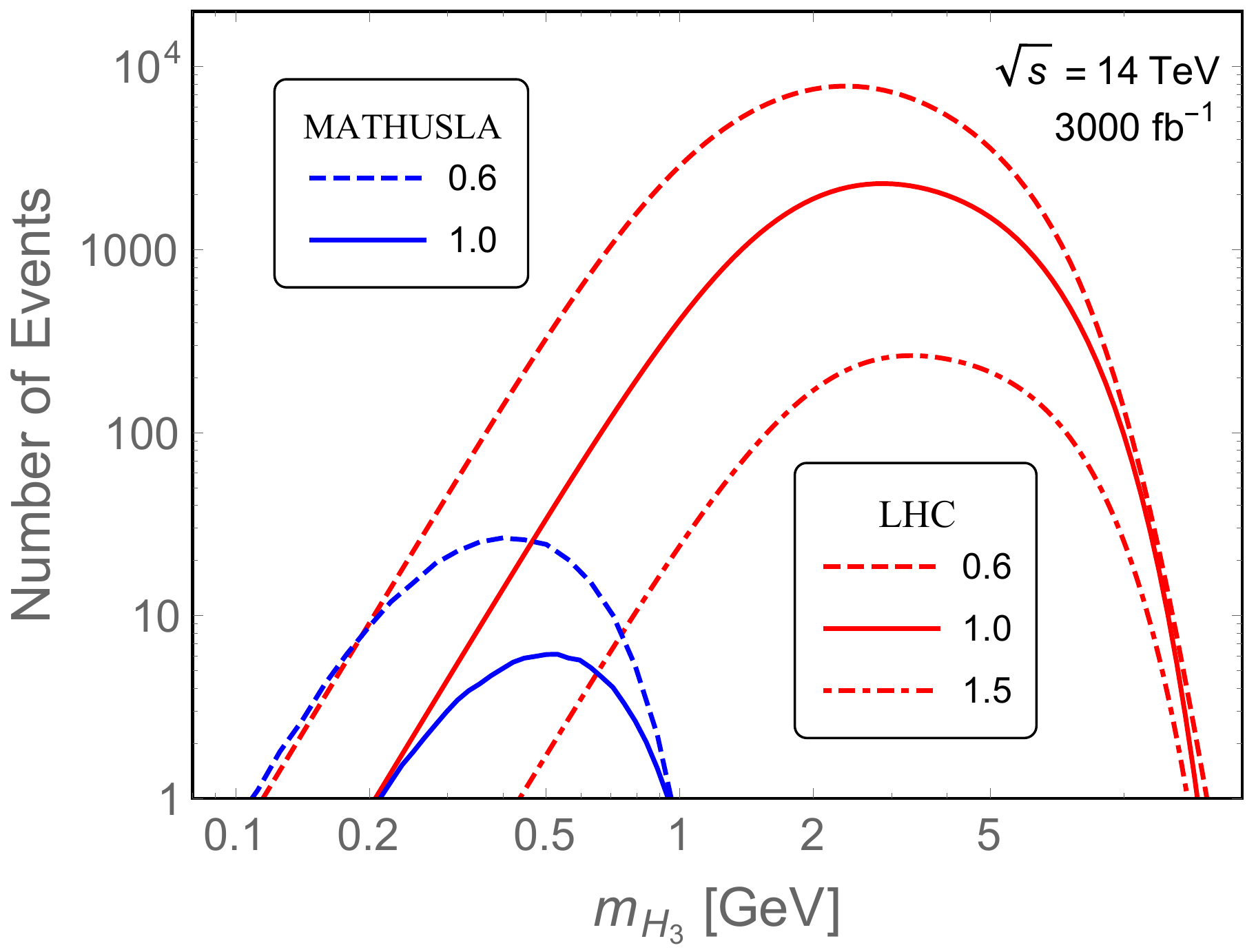}
  \caption{Predicted numbers of displaced photon events from $H_3$ decay within the ECAL of ATLAS and the surface detector MATHUSLA, with an integrated luminosity of 3000 fb$^{-1}$ at $\sqrt{s} = 14$ TeV, for $v_R = 5$ TeV and three benchmark values of $g_R / g_L = 0.6$, 1 and 1.5.}
  \label{fig:signal}
  \vspace{-10pt}
\end{figure}

Limited by the flavor data, a light $H_3$ decays mostly into the diphoton final state at the LHC after being produced. 
For a GeV mass, the decay-at-rest length $L_0$ is of order of cm; multiplied by a boost factor of $b\sim 100$, the actual decay length is expected to be of order of m, comparable to the radius of the Electromagnetic Calorimeter (ECAL) of ATLAS and CMS detectors, which are respectively 1.5 m~\cite{Aad:2009wy} and 1.3 m~\cite{Ball:2007zza}. The final-state photons from $H_3$ decay are highly collimated with a separation of $\Delta R \sim m_{H_3} / E_{H_3}$. Thus, most of the photon pairs can not be separated with the angular resolution of $\Delta \eta \times \Delta \phi = 0.025 \times 0.025$ (ATLAS) and $0.0174 \times 0.0174$ (CMS)~\cite{Aad:2009wy,Ball:2007zza}, and would be identified as a high-energy single-photon jet. Counting conservatively these single photon jets within $1\, {\rm cm} < L < R_{\rm ECAL}$, we can have up to thousands of signal events for an integrated luminosity of 3000 fb$^{-1}$ at $\sqrt{s} = 14$ TeV LHC, depending on the RH scale $v_R$ and gauge coupling $g_R$ (see Fig.~\ref{fig:signal}).
The SM fake rate for the displaced diphotons is expected to be small~\cite{Dasgupta:2016wxw}, thus the displaced photon events, with the associated VBF jets, would constitute a new ``smoking gun" signature of the $H_3$ decays as predicted by the minimal LRSM. For $m_{H_3}\lesssim 1$ GeV, the decay length exceeds the size of LHC detectors, but could be just suitable for future dedicated LLP search experiments, such as MATHUSLA~\cite{Chou:2016lxi}, as shown in Fig.~\ref{fig:limits}.

To have a better feeling of the displaced photon signal at the LHC and the dedicated long-lived particle surface detector MATHUSLA, we show here in Fig.~\ref{fig:signal} the expected numbers of signal events that could be collected in the ECAL of ATLAS and MATHUSLA, for the benchmark value of $v_R = 5$ TeV and $g_R / g_L = 0.6$, 1 and 1.5. The basic trigger cuts $p_T > 25$ GeV and $\Delta \phi_{jj} > 0.4$ are applied to the VBF jets. As the diphotons from $H_3$ decay are highly boosted, with a factor of $E_{H_3} / m_{H_3} \sim 10^2$, the photon pairs are highly collimated, and, to be conservative, we consider only the events that can {\it not} be separated by the ATLAS detector. At ATLAS, the displaced photon-jet signal could reach up to thousands; while at the surface detector MATHUSLA the effective solid angle is much smaller, $\lesssim 0.1 \times 4\pi$, thus the events are much less. However, far away from the collision point, ultra displaced signal at MATHUSLA is expected to be almost background-free. The LLP searches at the general-purpose detector ATLAS/CMS and dedicated detector MATHUSLA are largely complementary to each other.

\begin{figure}[!t]
  \centering
  \includegraphics[width=0.4\textwidth]{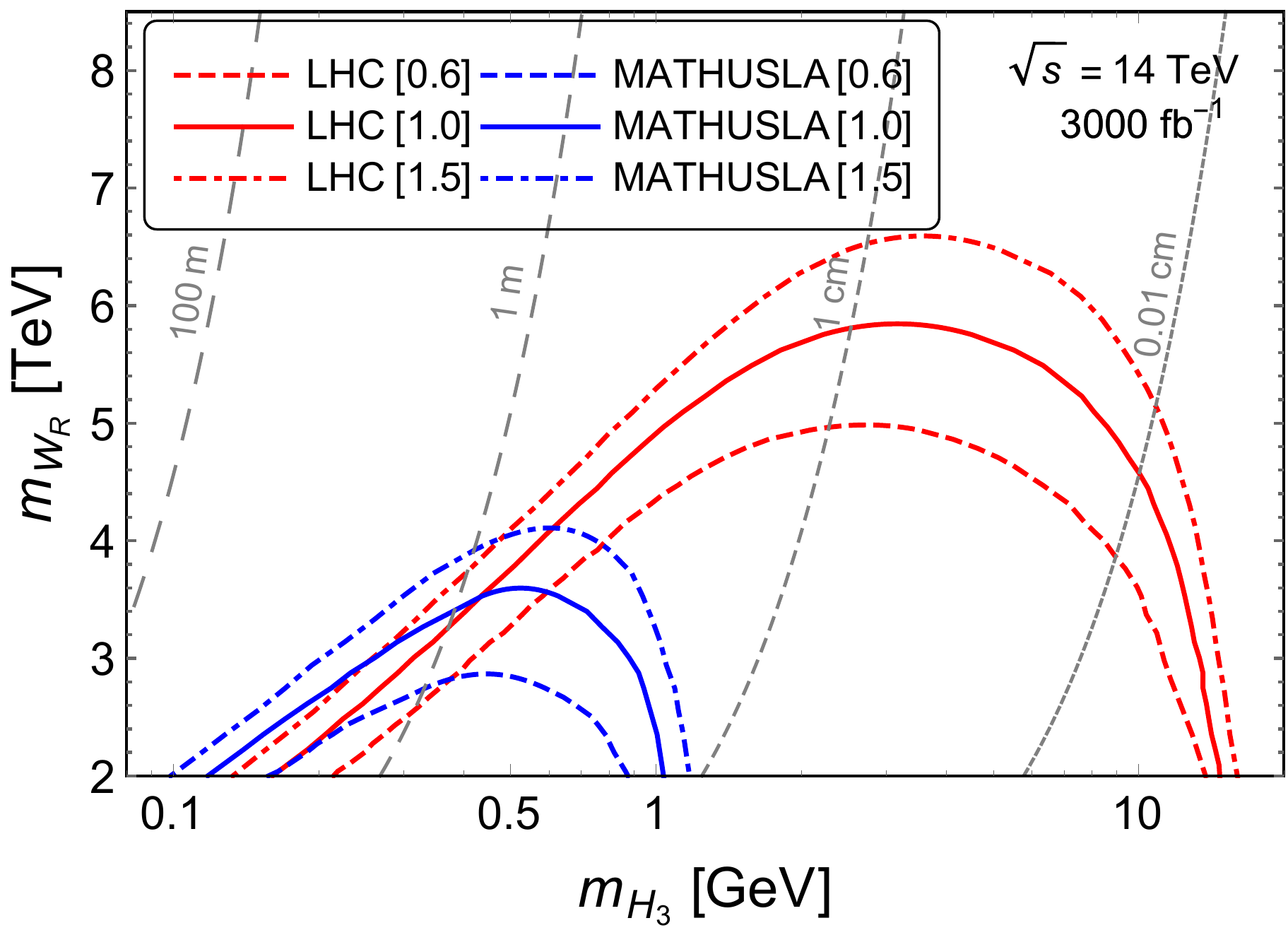}
  \caption{Sensitivity contours in the $m_{H_3} - m_{W_R}$ plane from LLP searches at LHC and MATHUSLA, for $g_R / g_L = 0.6$, 1 and 1.5. The grey contours indicate the proper lifetime of $H_3$ with $g_R = g_L$; for $g_R \neq g_L$, the lifetime has to be rescaled by the factor of $(g_R/g_L)^{-2}$.}
  \label{fig:LLP}
  \vspace{-10pt}
\end{figure}

The projected probable regions in the plane of $m_{H_3}$ and $m_{W_R}$ are presented in Fig.~\ref{fig:LLP}, for three benchmark values of $g_R / g_L = 0.6$, 1 and 1.5, where we have assumed 10 and 4 signal events of displaced photon jets at respectively LHC and MATHUSLA. As a result of the large Lorentz boost factors, the LLP searches at LHC and MATHUSLA are sensitive to larger values of $m_{H_3}$, as compared to the low-energy meson decay searches,  and are therefore complementary to the meson probes at the high intensity frontier, as clearly shown in Fig.~\ref{fig:limits}. This is also largely complementary to the direct searches of $W_R$ via same-sign dilepton plus jets in revealing the right-handed $SU(2)_R$ breaking and the TeV-scale seesaw mechanism at the high energy frontier, as shown in Fig.~\ref{fig:LLP}.

{\section{Summary}}
We have pointed out for the first time that, in the minimal LRSM the $SU(2)_R$ breaking scalar $H_3$ could be much lighter than the right-handed scale $v_R$, and searches for light $H_3$ via high energy displaced photon searches at the LHC provide a new probe of the TeV scale left-right seesaw models. We have derived the low energy flavor constraints on such particles, and given the predictions for the displaced photon signal from its production and decay at the LHC, as well as the prospects at the high-intensity frontier like SHiP and DUNE.  Moreover, the dominant diphoton decay channel of the light scalar considered here is a unique feature of the LRSM that can be used to distinguish it from other beyond SM scenarios.

\section*{Acknowledgement}
The work of R.N.M. is supported by the US National Science Foundation grant No.~PHY1620074. Y.Z. would like to thank the IISN and Belgian Science Policy (IAP VII/37) for support.

\begin{widetext}
\appendix*
\section{\bf Scalar potential, couplings and decay widths of $H_3$}
The most general renormalizable scalar potential for the $\Phi$ and $\Delta_R$ fields invariant under the gauge group ${\cal G}_{\rm LR}$ is given by
\begin{eqnarray}
\label{eqn:potential}
\mathcal{V} & \ = \ & - \mu_1^2 \: {\rm Tr} (\Phi^{\dag} \Phi) - \mu_2^2
\left[ {\rm Tr} (\tilde{\Phi} \Phi^{\dag}) + {\rm Tr} (\tilde{\Phi}^{\dag} \Phi) \right]
- \mu_3^2 \:  {\rm Tr} (\Delta_R
\Delta_R^{\dag}) 
+ \lambda_1 \left[ {\rm Tr} (\Phi^{\dag} \Phi) \right]^2 + \lambda_2 \left\{ \left[
{\rm Tr} (\tilde{\Phi} \Phi^{\dag}) \right]^2 + \left[ {\rm Tr}
(\tilde{\Phi}^{\dag} \Phi) \right]^2 \right\} \nonumber \\
&&+ \lambda_3 \: {\rm Tr} (\tilde{\Phi} \Phi^{\dag}) {\rm Tr} (\tilde{\Phi}^{\dag} \Phi) +
\lambda_4 \: {\rm Tr} (\Phi^{\dag} \Phi) \left[ {\rm Tr} (\tilde{\Phi} \Phi^{\dag}) + {\rm Tr}
(\tilde{\Phi}^{\dag} \Phi) \right]  
+ \rho_1  \left[ {\rm
Tr} (\Delta_R \Delta_R^{\dag}) \right]^2  
+ \rho_2 \: {\rm Tr} (\Delta_R
\Delta_R) {\rm Tr} (\Delta_R^{\dag} \Delta_R^{\dag}) \nonumber
\\
&&+ \alpha_1 \: {\rm Tr} (\Phi^{\dag} \Phi) {\rm Tr} (\Delta_R \Delta_R^{\dag})
+ \left[  \alpha_2 e^{i \delta_2}  {\rm Tr} (\tilde{\Phi}^{\dag} \Phi) {\rm Tr} (\Delta_R
\Delta_R^{\dag}) + {\rm H.c.} \right]
+ \alpha_3 \: {\rm
Tr}(\Phi^{\dag} \Phi \Delta_R \Delta_R^{\dag}) \,.
\end{eqnarray}
After symmetry breaking and diagonalization of the mass matrices, the physical scalar masses are
given by
\begin{align}
m_h^2 & \ \simeq \  \left(  4 \lambda _1-\frac{\alpha _1^2}{{\lambda_1-\rho_1}} \right) \kappa^2 \,, \label{eqn:hmass} \\
m_{H_1}^2 & \ \simeq \  \alpha _3 ( 1 + 2 \xi ^2 ) v_R^2 + 4 \left( 2 \lambda
   _2+\lambda _3 + \frac{4 \alpha _2^2 }{{\alpha _3}} \right) \kappa^2 \,,
\label{eqn:H10mass} \\
m_{H_3}^2 & \ \simeq \ 4 \rho_1 v_R^2 + \left( \frac{\alpha_1^2}{\lambda_1-\rho_1} - \frac{16 \alpha_2^2}{{\alpha_3}} \right) \kappa^2 \,, \label{eqn:H30mass} \\
m_{A_1}^2 & \ \simeq \ \alpha _3 ( 1 + 2 \xi ^2 ) v_R^2 +4 \left(\lambda _3-2 \lambda _2\right) \kappa^2 \,, \label{eqn:A10mass} \\
m^2_{H_1^\pm} & \ \simeq \ \alpha _3 \left[(1 +2 \xi ^2) v_R^2 + \frac12 \kappa^2\right] \,, \label{eqn:H1pmass} \\
m^2_{H_2^{\pm\pm}} & \ \simeq \ 4 \rho_2 v_R^2 + \alpha_3 \kappa^2 \,, \label{eqn:H1ppmass}
\end{align}
where $\xi\equiv \kappa'/\kappa$ is the ratio of the bidoublet VEVs.

All the couplings of $H_3$ to the SM and heavy particles in the LRSM are given in Table~\ref{tab:coupling}, which is based on the calculation of Ref.~[7] and up to the leading order in the small parameters $\xi$, $\epsilon\equiv v_{\rm EW}/v_R$, $\sin\tilde\theta_1 = \sin\theta_1 + \xi \sin\theta_2$,  $\sin\tilde\theta_2 = \sin\theta_2 + \xi \sin\theta_1$. Here $\phi$ is defined as $\tan\phi \equiv g_{BL} / g_{R}$.
\begin{table}[!h]
  \centering
  \caption[]{The couplings of a light scalar $H_3$. The mixing angles $\theta_1$ and $\theta_2$ are defined in Eq.~(3).}
  \label{tab:coupling}
  \begin{tabular}[t]{lll}
  \hline\hline
  couplings & values \\ \hline
  $H_3 hh$ & $ \frac{1}{\sqrt{2}} \alpha _1 v_R $ \\
  $h H_3 H_3$ & $ - \sqrt2 \alpha _1 v_{\rm EW} $ \\
  $H_3 h H_1$ & $ 2 \sqrt{2} \alpha _2 v_R $ \\
  $H_3 H_1 H_1$ & $ \frac{1}{\sqrt{2}} \alpha _3 v_R $  \\
  $H_3 A_1 A_1$ & $\frac{1}{\sqrt{2}} \alpha _3  v_R$  \\
  $H_3 H_1^+ H_1^-$ & $\sqrt{2} \alpha _3  v_R$  \\
  $H_3 H_2^{++} H_2^{--}$ & $2 \sqrt{2} \left(\rho _1+2 \rho _2\right) v_R $ \\ \hline
  $H_3 \bar{u} u$ & $\frac{1}{\sqrt2} \widehat{Y}_U
  \sin\tilde\theta_1
  - \frac{1}{\sqrt2} \left( V_L \widehat{Y}_D V_R^\dagger \right) \sin\tilde\theta_2 $\\
  $H_3 \bar{d}d$  & $\frac{1}{\sqrt2} \widehat{Y}_D
  \sin\tilde\theta_1
  -\frac{1}{\sqrt2} \left( V_L^\dagger \widehat{Y}_U V_R \right) \sin\tilde\theta_2$\\
  $H_3 \bar{e}e$  & $\frac{1}{\sqrt2} \widehat{Y}_E
  \sin\tilde\theta_1 -\frac{1}{\sqrt2} Y_{\nu N}
  \sin\tilde\theta_2$ \\
  $H_3 NN$  & $\frac{M_N}{\sqrt2 v_R}$ \\
  \hline\hline
  \end{tabular}
\hspace{0.2cm}
  \begin{tabular}[t]{lll}
  \hline\hline
  couplings & values \\ \hline
  $H_3  W_{}^+ W_{}^{-}$ & $\frac{1}{\sqrt2}g_L^2 \sin\theta_1\, v_{\rm EW} + \sqrt2 g_R^2 \sin^2\zeta_W \, v_R$ \\
  $H_3  W_{}^+ W_R^{-}$ & $\sqrt2 g_R^2 \sin\zeta_W \, v_R$ \\
  $H_3  W_{R}^+ W_R^{-}$ & $\sqrt2  g_R^2 v_R $ \\ \hline
  $H_3  Z_{} Z^{}$ & $\frac{g_L^2 \sin\theta_1\, v_{\rm EW}}{2\sqrt2 \cos^2\theta_W} +
  \frac{\sqrt2 g_R^2 \sin^2\zeta_Z \, v_R}{\cos^2\phi} $ \\
  $H_3  Z_{} Z_{R}$ & $-\frac{g_L g_R \sin\theta_1 \cos\phi \, v_{\rm EW}}{\sqrt2 \cos\theta_W} +
  \frac{2\sqrt2 g_R^2 \sin\zeta_Z \, v_R }{\cos^2\phi} $ \\
  $H_3  Z_{R} Z_R^{}$ & $\frac{\sqrt2 g_R^2 v_R }{\cos^2\phi} $ \\ \hline
  $H_3 H_1^+ W^-$ & $\frac12 g_L (\sin\theta_2 - \sin\theta_1 \xi)$ \\
  $H_3 H_1^+ W_R^-$ & $\frac12 g_R \epsilon$ \\
  $H_3 A_1 Z$ & $-\frac{i g_L (\sin\theta_2 - \sin\theta_1 \xi)}{2\cos\theta_W}$ \\
  $H_3 A_1 Z_R$ & $\frac{i}{2} g_R (\sin\theta_2 - \sin\theta_1 \xi) \cos\phi$ \\
  \hline\hline
  \end{tabular}
\end{table}

The partial decay widths for the dominant decay modes of $H_3$ are collected below:
\begin{eqnarray}
\Gamma (H_3 \to q\bar{q}) &=&
\frac{3 m_{H_3}}{16\pi}
\left[ \sum_{i,j} \left| \mathcal{Y}_{u, \, ij} \right|^2
\beta^3_2 (m_{H_3}, m_{u_i}, m_{u_j})
\Theta (m_{H_3}- m_{u_i}- m_{u_j}) \nonumber \right.  \\
&& \qquad\quad + \left. \sum_{i,j} \left| \mathcal{Y}_{d, \, ij} \right|^2
\beta^3_2 (m_{H_3}, m_{d_i}, m_{d_j})
\Theta (m_{H_3}- m_{d_i}- m_{d_j})  \right] , \\
\Gamma (H_3 \to \ell^+ \ell^-) &=&
\frac{m_{H_3}}{16\pi}
\sum_{i,j} \left| \mathcal{Y}_{e, \, ij} \right|^2
\beta^3_2 (m_{H_3}, m_{e_i}, m_{e_j})
\Theta (m_{H_3}- m_{e_i}- m_{e_j}) \,, \\
\Gamma (H_3 \to \gamma\gamma) &=&
\frac{\alpha^2 m_{H_3}^3}{1028 \pi^3}
\left| \frac{\sqrt2}{v_R} A_{0}(\tau_{H_1^\pm}) +
\frac{4\sqrt2}{v_R} A_{0}(\tau_{H_2^{\pm\pm}}) +
\frac{\sqrt2}{v_{\rm EW}} \sum_{f = q, \ell} f_f N^f_C Q_f A_{1/2} (\tau_f) +
\frac{\sqrt2}{v_R} A_{1} (\tau_{W_R}) \right|^2 \,, \label{eqn:gamma} \\
\Gamma (H_3 \to gg) &=&
\frac{G_F \alpha_s^2 m_{H_3}^3}{36 \sqrt2 \pi^3}
\left| \frac34 \sum_{f = q} f_f A_{1/2} (\tau_f) \right|^2 \,,
\end{eqnarray}
with the kinetic function
\begin{eqnarray}
\label{eqn:beta2}
\beta_2 (M,\, m_1,\, m_2) & \ \equiv \ & \left[ 1 - \frac{2(m_1^2 + m_2^2)}{M^2} + \frac{(m_1^2 - m_2^2)^2}{M^4} \right]^{1/2} \,,
\end{eqnarray}
the Yukawa couplings
\begin{align}
\mathcal{Y}_{u} & \ = \ \widehat{Y}_U
  \sin\tilde\theta_1
  -  \left( V_L \widehat{Y}_D V_R^\dagger \right) \sin\tilde\theta_2,\\
\mathcal{Y}_{d} & \ = \ \widehat{Y}_D
  \sin\tilde\theta_1
  -\left( V_L^\dagger \widehat{Y}_U V_R \right) \sin\tilde\theta_2,\\
\mathcal{Y}_{e} & \ = \ \widehat{Y}_E
  \sin\tilde\theta_1 - Y_{\nu N}
  \sin\tilde\theta_2 \, ,
\end{align}
$f_f$ the normalization factor with respect to the SM Yukawa couplings,
\begin{eqnarray}
f_{u,i} &=& \sin\tilde\theta_1
- \frac{(V_L \widehat{M}_d V_R^\dagger)_{ii}}{m_{u,i}}
\sin\tilde\theta_2 \,, \\
f_{d,i} &=& \sin\tilde\theta_1
- \frac{(V_L^\dagger \widehat{M}_u V_R)_{ii}}{m_{d,i}}
\sin\tilde\theta_2 \,, \\
f_{e,i} &=& \sin\tilde\theta_1
- \frac{Y_{\nu N, ii}}{m_{e,i}/v_{\rm EW}}
\sin\tilde\theta_2 \,,
\end{eqnarray}
and the loop functions
\begin{eqnarray}
A_{0} (\tau) &\equiv& - \left[ \tau - f(\tau) \right] \tau^{-2} \,, \\
A_{1/2} (\tau) &\equiv& 2 \left[ \tau + (\tau-1) f(\tau) \right] \tau^{-2} \,, \\
A_1 (\tau) &\equiv& - \left[ 2 \tau^2 + 3\tau + 3 (2\tau-1) f(\tau) \right] \tau^{-2} \,,
\end{eqnarray}
with $\tau_X = m_{H_3}^2 / 4m_X^2$ and
\begin{align}
f(\tau) \ \equiv \ \left\{ \begin{array}{cc}
{\rm \arcsin}^2\sqrt{\tau} & ({\rm for}~\tau\leq 1) \\
-{\displaystyle \frac{1}{4}}\left[\log \left( \frac{1+\sqrt{1-1/\tau}}{1-\sqrt{1-1/\tau}}\right)-i\pi  \right]^2 & ({\rm for}~\tau>1) \;.
\end{array}\right.
\label{fx}
\end{align}
For the heavy particle loops, only the large loop mass limit is useful for us: $A_0 (0) = 1/3,\, A_{1/2} (0) = 4/3, \, A_1 (0) = -7$.
In this limit the gauge decay mode $\gamma\gamma$ is only sensitive to the RH scale $v_R$ via $\Gamma \propto v_R^{-2}$. The contributions from the scalars $H_1^\pm$ and $H_2^{\pm\pm}$are suppressed by $5A_0 (0) / A_{1} (0) = -5/21$, with the factor of 5 from sum of the electric charges squared.

\end{widetext}


\end{document}